\documentclass[onecolumn,doublespacing, superscriptaddress]{revtex4}
\usepackage{graphicx}
\usepackage{dcolumn}
\usepackage{bm}
\usepackage{amsmath}
\usepackage{epsfig}
\begin{document}

\title{Neutron response of PARIS phoswich detector}

\author{Balaram Dey}
\affiliation{Department of Nuclear and Atomic Physics, Tata Institute of Fundamental Research, Mumbai-400005, India}
\author{C. Ghosh}
\affiliation{Department of Nuclear and Atomic Physics, Tata Institute of Fundamental Research, Mumbai-400005, India}
\author{S. Pal}
\affiliation{Pelletron Linac Facility, Tata Institute of Fundamental Research, Mumbai-400005, India}
\author{V. Nanal}
\email[e-mail:]{nanal@tifr.res.in}
\affiliation{Department of Nuclear and Atomic Physics, Tata Institute of Fundamental Research, Mumbai-400005, India}
\author{R.G. Pillay}
\affiliation{Department of Nuclear and Atomic Physics, Tata Institute of Fundamental Research, Mumbai-400005, India}
\author{K.V. Anoop}
\affiliation{Pelletron Linac Facility, Tata Institute of Fundamental Research, Mumbai-400005, India}
\author{M.S. Pose}
\affiliation{Department of Nuclear and Atomic Physics, Tata Institute of Fundamental Research, Mumbai-400005, India}


\date{\today}

\begin{abstract}
We have studied neutron response of PARIS phoswich [LaBr$_3$(Ce)-NaI(Tl)] detector which is being developed for measuring the high energy (E$_{\gamma}$ = 5 - 30 MeV) $\gamma$ rays emitted from the decay of highly collective states in atomic nuclei. The relative neutron detection efficiency of LaBr$_3$(Ce) and NaI(Tl) crystal of the phoswich detector has been measured using the time-of-flight (TOF) and pulse shape discrimination (PSD) technique in the energy range of E$_n$ = 1 - 9 MeV and compared with the GEANT4 based simulations. It has been found that for E$_n$ $>$ 3 MeV, $\sim$ 95 \% of neutrons have the primary interaction in the LaBr$_3$(Ce) crystal, indicating that a clear n-$\gamma$ separation can be achieved even at $\sim$15 cm flight path.
\end{abstract}
\pacs{24.30.Cz,24.60.Dr,25.70.Gh}
\maketitle
\section{Introduction}
\label{sec:1}

The measurement of high energy $\gamma$-rays, emitted from the decay of highly collective states of atomic nuclei \cite{harakeh} as well as from nucleon-nucleon bremsstrahlung during the early stages of the target-projectile collision \cite{nif}, is an excellent probe to study the atomic nuclei under the extreme conditions of temperature and angular momentum. Earlier, the study of high energy $\gamma$-rays have been performed by many groups in the world and their studies are confined to regions near the valley of stability \cite{bdey, ghosh1, bracco}.
In the last few years, it is becoming possible to investigate the unexplored regions of the nuclear chart (especially on the neutron-rich side) with the availability of beams of short-lived nuclei i.e. radio-active ion beams (RIB).  

A Photon Array for the Studies with Radioactive Ion and Stable beams (PARIS) is being developed in order to measure the high energy $\gamma$-rays (E$_{\gamma}$ = 5 - 30 MeV) \cite{paris,adam}.
The array consists of $\sim$200 PARIS phoswich element. Each element is made up of 2$^{\prime\prime}\times$ 2$^{\prime\prime}\times$2$^{\prime\prime}$ LaBr$_3$(Ce) crystal optically coupled to a 2$^{\prime\prime}\times$ 2$^{\prime\prime}\times$6$^{\prime\prime}$ NaI(Tl) crystal followed by a single PMT for signal readout. Recently, two PARIS phoswich detector element have been characterized over a wide range of $\gamma$-ray energies \cite{ghosh2}.
In the high energy $\gamma$-ray measurement, the evaporated neutrons are the major sources of contamination, which can be rejected by time-of-flight (TOF) technique. On the other hand, the array should be placed as much close as possible to the target chamber due to low intensity radio-active ion beam, in order to enhance the efficiency of the array which is important for low cross-section measurement. But, the rejection of neutron background becomes difficult if the detector array is placed very close to the target chamber.
Therefore, the study of neutron response of the phoswich detector is very essential for the rejection of neutron contamination in the high $\gamma$- ray measurement as well as to optimize the distance between detector and center of the target chamber.

One of the major advantages of LaBr$_3$ in the PARIS design is the excellent timing characteristics, enabling n-$\gamma$ discrimination using time-of-flight (TOF) technique at close distances from target, which is essential for low intensity RIB experiments ~\cite{paris,adam}. 

\begin{figure*}[htp]
\centering
\includegraphics[scale=2.0]{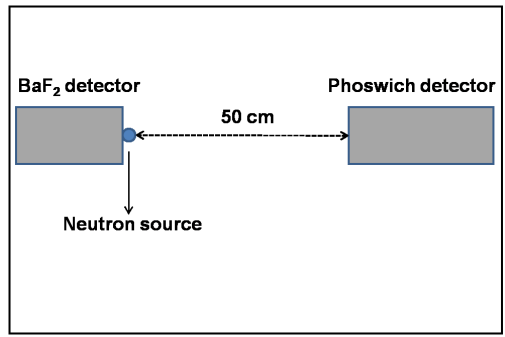}  
\caption{A schematic diagram of experimental set-up.}
\label{exp_setup} 
\end{figure*}

The n-$\gamma$ separation for smaller flight paths relies on the neutron interaction in LaBr$_3$(Ce) crystal of the PARIS phoswich detector.
It is known that the neutrons of energy E$_n$ $<$ 10 MeV interact mainly through (n,$\gamma$) or (n, n$^{\prime}\gamma$) reactions, while E$_n$ $>$ 10 MeV more complicated reactions are involved producing charged hadrons \cite{tain}.
The interaction probability is high in the LaBr$_3$ crystal of phoswich detector owing to its high density (5.08 gm/cm$^3$). 
It is therefore important to measure the relative neutron detection efficiency in LaBr$_3$(Ce) and NaI(Tl) crystals of the PARIS-phoswich detector. 

\begin{figure*}[htp]
\centering
\includegraphics[scale=0.48]{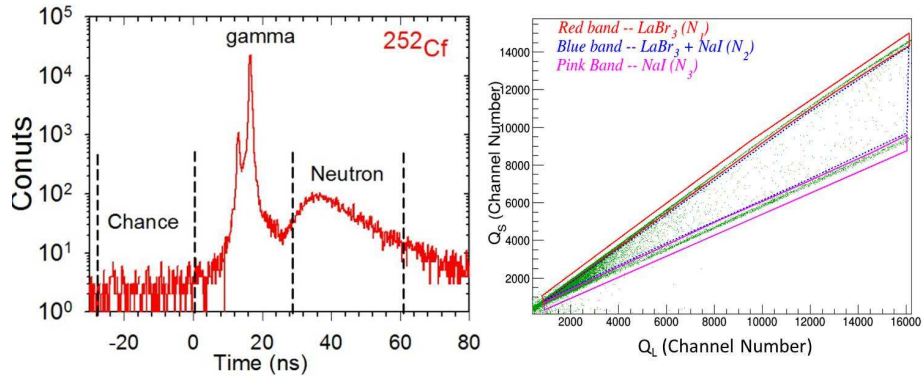} 
\caption{Time-of-flight spectrum (left panel) and PSD spectrum (right panel) using $^{252}$Cf source.  }
\label{tof} 
\end{figure*}

In this work, we have measured the relative neutron detection efficiency of LaBr$_3$(Ce) and NaI(Tl) crystal of the phoswich detector as a function of neutron energy (1-9 MeV) and compared with the GEANT4 based simulation \cite{geant4}. 
\vspace{-7mm}

\section{Experimental details}
\label{sec:2}
\vspace{-4mm}
The measurements were carried out at TIFR, Mumbai using different neutron sources $^{252}$Cf, $^{241}$Am-$^{9}$Be and $^{239}$Pu-$^{13}$C. A schematic diagram of experimental set-up is shown in Fig. \ref{exp_setup}. 
A phoswich detector was placed at 50 cm distance from the source and a BaF$_2$ detector (hexagonal, 9 cm long and 6 cm face-to-face) was placed close to the source ($<$1 cm). 
The TOF technique was employed to measure the neutrons, for which the START and STOP triggers were taken from the BaF$_2$ and the phoswich detector, respectively.
A CAEN make VME based digitizer V1751 (1 GHz, 1 Vpp, 10 bit) was used to acquire the data. Timing information was extracted using an algorithm implementing constant fraction discrimination with a delay of 6 ns and 20\% fraction, incorporated in the online WAVEDUMP acquisition software \cite{ghosh1,anoop}. 
The TOF spectra for two different neutron sources are shown in the left panel of Fig. \ref{tof}. A small peak to the left of the $\gamma$-prompt peak, corresponds to neutron events in BaF$_2$ and $\gamma$ events in the phoswich detector. This was confirmed by varying the neutron flight path. The time window used for chance correction is shown in the left panel of Fig. \ref{tof}. 
The data was also recorded without any neutron source to assess the background, which was found to be negligibly small.

Neutron energies are calculated from the neutron TOF using prompt $\gamma$-peak as a time reference.
The identification of LaBr$_3$/NaI events was done by the pulse shape discrimination. The output pulse from photo multiplier tube (PMT) was integrated and recorded for different gate widths of 300 and 900 ns, to get the Q$_S$ (corresponding to energy deposition in LaBr$_3$) and Q$_L$ (corresponding to the energy deposition in NaI), respectively.
The neutron TOF gated 2D spectrum of Q$_L$-Q$_S$ is shown in the right panel of Fig. \ref{tof}.

\begin{figure*}[htp]
\centering
\includegraphics[scale=0.35]{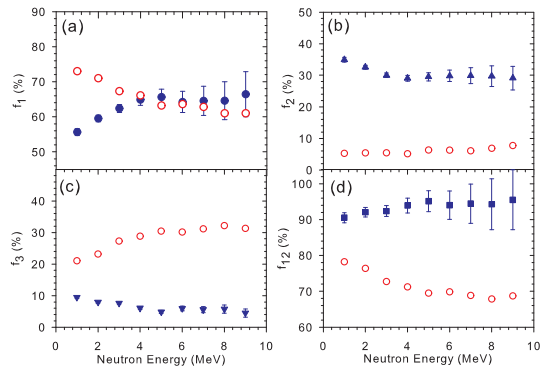} 
\caption{Relative detection efficiency of neutrons in the phoswich detector (See text for details). Filled symbols represent the experimental data ($^{252}$Cf) and simulations are shown by open symbols.}
\label{fig1} 
\end{figure*}

\section{Relative neutron detection efficiency}
\vspace{-4mm}
The observed neutron events in the phoswich detector are classified into three categories corresponding to energy deposition as shown in the righ panel of Fig. \ref{tof} : energy in only LaBr$_3$ (N$_1$), energy in both the LaBr$_3$ and NaI (N$_2$), and energy in only NaI (N$_3$). 
The relative neutron detection (f$_i$), defined as N$_i$/N$_{tot}$ (N$_{tot}$ = N$_1$ + N$_2$ + N$_3$), is computed as a function of neutron energy for all three categories as shown in Fig ~\ref{fig1}. 
In addition, f$_{12}$ = (N$_1$ + N$_2$)/N$_{tot}$, corresponding to primary neutron interaction in LaBr$_3$, is also shown in Fig. \ref{fig1}(d).
The data for different neutron sources were found to be consistent within the error bars.
The present results have been compared with the GEANT4 sumilations as shown in Fig. \ref{fig1} (Open symbols). The neutron data library ENDF/B-VI was used in the simulation code.
\vspace{-7mm}

\section{Results and discussions}
\label{Sec:4}
\vspace{-4mm}

As can be seen in Fig. \ref{fig1}(a),  about 60 - 70 \% of neutron energy deposition (N$_1$/N$_{tot}$) take place in only the LaBr$_3$ crystal of the phoswich detector. It is evident that the neutron interaction probability in LaBr$_3$ increases from 1 to 3 MeV and is nearly constant between 3 - 9 MeV.  
The simulations reproduce the observed trend at E $>$ 4 MeV, but values are lower by about 10$\%$ than the data. However, the simulation result underpredicts the data for primary interaction in LaBr$_3$ (Fig. \ref{fig1}b), whereas it overpredicts the data for primary interaction in NaI (Fig. \ref{fig1}c) and hence mixed events are mismatched with the data (Fig. \ref{fig1}d).
This discrepancy could be due to discrepancies in the neutron libraries used in the simulation, which was also reported in Ref.~\cite{tain}. 
Experimentally, it is clearly observed that $\sim$90 - 95 \% of neutron events have the primary interaction in LaBr$_3$ crystal of the phoswich detector as can be seen from Fig. \ref{fig1}(d). It should be mentioned that mixed events could also have contribution from primary interaction in NaI and subsequent back-scattering of neutron, $\gamma$, e$^-$/e$^+$ in the LaBr$_3$. This contribution was estimated from simulation and found to be $\sim$1 - 3 \% of total events in the phoswich detector. 
\vspace{-7mm}

\section{Summary and Conclusion}
\label{Sec:4}
\vspace{-4mm}
The response of a LaBr$_3$(Ce)-NaI(Tl) phoswich detector to low energy neutrons (E$_n$ $<$ 10 MeV) have been measured using different neutron sources and compared with the GEANT4 based simulation. The present study has shown that $\sim$90 - 95\% of neutrons with E $>$ 3 MeV have primary interaction in the LaBr$_3$ part of the PARIS phoswich detector. Thus, a clear n-$\gamma$ separation can be achieved even at $\sim$ 15 cm flight path for majority of events. For energies below 3 MeV, the flight time is sufficiently large at $\sim$ 15 cm distance (T $>$ 6 ns; enabling the n-$\gamma$ separation) even in case of primary neutron interactions in NaI. Hence, overall neutron rejection with $90\%$ probability is feasible with flight paths of $\sim$15 cm in the phoswich detector. 

\vspace{-7mm}

\section{Acknowledgments}
\vspace{-4mm}
We would like to thank Mr. K.S. Divekar for assistance with setup. We would also like to thank all the PARIS collaborators.

\end{document}